# Analog-to-digital Conversion Revolutionized by Deep Learning

Shaofu Xu[1], Xiuting Zou[1], Bowen Ma[1], Jianping Chen[1], Lei Yu[1], Weiwen Zou[1,*]

[1]State Key Laboratory of Advanced Optical Communication Systems and Networks, Department of Electronic Engineering, Shanghai Jiao Tong University, Shanghai 200240, China.

*Correspondence to: wzou@sjtu.edu.cn.

**Abstract:** As the bridge between the analog world and digital computers, analog-to-digital converters are generally used in modern information systems such as radar, surveillance, and communications. For the configuration of analog-to-digital converters in future high-frequency broadband systems, we introduce a revolutionary architecture that adopts deep learning technology to overcome tradeoffs between bandwidth, sampling rate, and accuracy. A photonic front-end provides broadband capability for direct sampling and speed multiplication. Trained deep neural networks learn the patterns of system defects, maintaining high accuracy of quantized data in a succinct and adaptive manner. Based on numerical and experimental demonstrations, we show that the proposed architecture outperforms state-of-the-art analog-to-digital converters, confirming the potential of our approach in future analog-to-digital converter design and performance enhancement of future information systems.

From the advent of digital processing and the von Neumann computing scheme, the continuous world has become discrete by use of analog-to-digital converters (ADCs). Discrete digital signals are easier to process, store, and display; thus, they are integral to modern electronic information systems such as radar, surveillance, and communication systems. Following the major trend of performance enhancement, next-generation information systems are aiming at achieving high operating frequencies and broad bandwidths (1–3). Under these increasing demands, ADCs should be developed toward achieving high sampling rates, broad bandwidths, and sufficient quantization accuracy. Although electronic technologies conduct analog-to-digital conversion in most contemporary information systems and exhibit high accuracy owing to excellent manufacturing of electronic components, they suffer heavily from the bottleneck of electronic timing jitter and exhibit inferior performance in terms of high-speed sampling (4, 5). Additionally, the difficulty in manufacturing broadband electronic components hinders their high-frequency and broad-bandwidth applications (3, 6). As an elegant approach to facilitate the further development of ADCs, photonic technologies provide lower noise and broadband capability with ultra-low timing jitter, broadband radio frequency (RF) direct sampling, and ultra-high sampling rates (4). However, imperfect setup of photonic components gives rise to system defects and can deteriorate the performance of ADCs. As an essential component of photonic ADC architecture, electrooptic modulators typically suffer from nonlinear transmission function and consequently the linear dynamic range is severely limited (4). Although balanced detection methods (7) have been proposed to eliminate the nonlinearity, they still require complicated setups and miscellaneous data processing steps. Additionally, imperfect optical multichannelization results in mismatched distortions and can adversely affect quantization accuracy (8, 9). State-of-the-art mismatch compensation algorithms are based on frequency-domain analysis (9), which is inappropriate for aliased spectra and is difficult to accelerate. Currently, overcoming the tradeoff between bandwidth, sampling rate, and accuracy (dynamic range) remains a challenge for all existing ADC architectures.

Deep learning techniques involve a family of data processing algorithms that use deep neural networks to manipulate data (10). Deep learning has made significant advances in a variety of artificial intelligence applications such as computer vision (11, 12), medical diagnosis (13), and gaming (14). By constructing multiple layers of neurons and applying appropriate training methods, data from images, audio, and video can be automatically extracted with representations to be used in the inference of unknown data. The performance of deep learning algorithms surpasses that of human beings in several areas. Particularly, data recovery tasks including speech enhancement (15), image denoising (16), and reconstruction (17, 18) are well accomplished with convolutional neural networks (CNNs, neural networks based on convolutional filters), confirming the ability of deep neural networks to learn the model of data contamination and distortion, as well as output the recovered data.

We present a deep-learning-powered analog-to-digital conversion architecture that overcomes the electronic bottlenecks of timing jitter, bandwidth limitation, and photonic defects of nonlinearity and channel mismatch, simultaneously exploiting the advantages of electronic and photonic technologies. Consequently, this architecture features broad bandwidth, high speed, and high accuracy. Based on this architecture, we set up a two-channel 20 giga-samples per second (GS/s) ADC for experimental demonstration. The results show that the deep learning method is effective in eliminating the distortions of different signal formats, and that the ADC setup performs well compared with state-of-the-art ADCs. Furthermore, we conduct a series of simulations to demonstrate system expandability for greater number of channels and a supplementary experiment to demonstrate the attainable ultrahigh accuracy and dynamic range. The results confirm that the proposed architecture revolutionizes analog-to-digital conversion with high accuracy and high sampling rate in broadband. The architecture is also easily trainable and expandable,

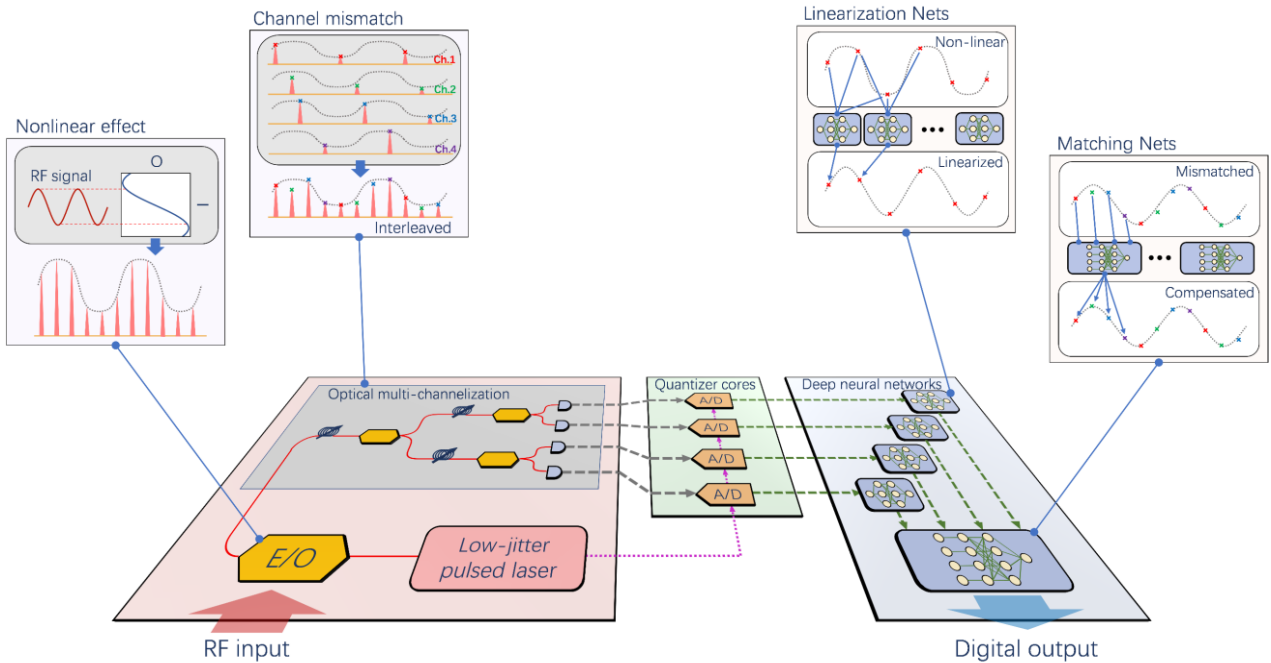

**Fig. 1. Schematic representations of deep-learning-powered analog-to-digital conversion architecture.** Three cascaded parts (photonic front-end, electronic quantization, and deep learning data recovery) are illustrated with different background colors. In the photonic front-end, E/O provides a broad bandwidth for receiving RF but causes nonlinearity to the system because of a sinusoidal-like transfer function. Illustrated in the subplot is the nonlinear effect: a standard sine RF signal that will be distorted by the nonlinear transfer function, distorting the sampling pulses. Besides, to our knowledge, all the reported methods of multichannelization (a two-stage optical time-divided demultiplexing method is depicted as an example) introduce channel mismatches (see subplot of channel mismatch). To ensure the high accuracy of electronic quantizers, not only should the low-jitter pulsed laser offer a high-quality clock, but the distorted signal should also be recovered. Cascaded after each electronic quantizer, the neural network linearization nets recover the nonlinearity distortions by convolution; the matching nets receive all channels' linearized output data and interleave them with the mismatch compensated.

ensuring its potential use in future broadband and highly dynamic information system applications.

**The analog-to-digital conversion architecture**

The deep-learning-powered analog-to-digital conversion architecture is mainly composed of three cascaded parts (Fig. 1): photonic front-end (4), electronic quantization, and deep learning data recovery. In the photonic front-end, a low-jitter pulsed laser source (8, 9) provides the sampling optical pulse train and the quantization clock. The pulses in the pulse train have a fixed repetition rate that can be regarded as the overall sampling rate of the analog-to-digital conversion. Subsequently, the optical pulse train samples the input RF signals in the electrooptic modulator (E/O). Typically, E/O conducts intensity modulation; hence, the intensity of each optical pulse is a sample of the input RF. A major advantage of an E/O is broadband reception, allowing high-frequency RF to be directly input and sampling to be performed without multilevel down-conversion that is generally used in conventional RF receiving schemes; therefore, noise accumulation is avoided. Since the repetition rate of the optical pulses is too high for an electronic quantizer that is relatively low-speed, multichannelization should be performed in the photonic front-end. Dominant schemes include time-to-wavelength demultiplexing (8, 9) and time-division demultiplexing (19). Figure 1 shows a schematic of four-channel time-division demultiplexing as an example. In each stage of demultiplexing, two adjacent optical pulses are divided into two output channels. After $N$ stages of demultiplexing, the repetition rate of the optical pulse train in each channel is divided by $2^N$, compatible with the electronic quantizers. At the end of the photonic front-end, photo detectors (PDs) convert the multichannelized optical pulse train to electrical signals for the next part of electronic quantization. Synchronized with the clock generated by the low-jitter pulsed laser, electronic quantizers (A/D) convert the electrical signals to digital data. Given their lowered sampling rate, electronic quantizers can offer high accuracy. However, the accuracy of holistic analog-to-digital conversion is severely limited by system defects. Two major defects induced by the photonic front-end are illustrated in Fig. 1, one of which is the nonlinear effect of the E/O. Theoretically, since the intensity modulation effect is a consequence of internal phase shift and optical interference, the transfer function is sinusoidal-alike. Additionally, in practice, the transfer function can be shifted by an inappropriate bias voltage and manufacturing imperfections. Nonlinearity in the E/O will greatly limit the dynamic range of the system, and thus degrade the accuracy of analog-to-digital conversion. Besides, another defect induced by the photonic front-end is channel mismatch, which is caused by mismatched delays, attenuations, and even sampling pulse shapes of different channels. These mismatches will lead to large distortions when the channels are interleaved to one. To overcome the effects of these system defects and maintain high accuracy, deep learning data recovery is deployed at the end of analog-to-digital conversion. For nonlinearity correction, deep neural networks named 'linearization nets' are tailed at each channel. After training, these neural networks will calculate the nonlinearity eliminated from the original data. Linearized data from all channels then enter other deep neural networks that we call 'matching nets', which perform matched interleaving: the matching nets will compensate for the mismatches after training.

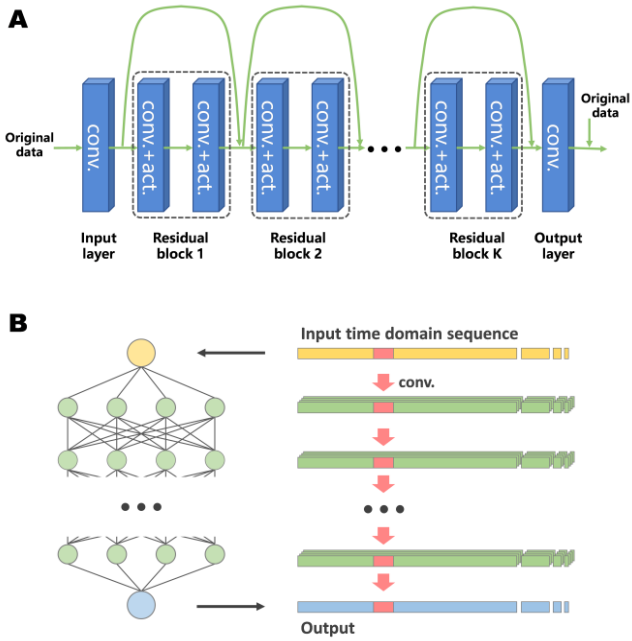

**Fig. 2. Basic model of linearization nets and matching nets.** (A) The adopted residual-on-residual learning model. conv. represents convolution and act. denotes nonlinear activation. Note that the output should be added up with the original input data. (B) Schematic of data manipulation in purely convolutional neural networks where all links are convolutions. Every convolutional window has a confined actuating range (red block) and they yield results by moving the window in sequence.

**Experimental implementation**

Experimental demonstrations of the high-speed high-accuracy ADCs were set up to test the validity of the proposed analog-to-digital conversion architecture. Mode-locked lasers (MLLs) were adopted as the low-jitter pulsed laser source, which provides high-repetition-rate optical pulse train for the system. To realize the direct sampling of high-frequency RF signals, we implemented the E/O with a Mach-Zehnder modulator, which has a bandwidth up to 40 GHz. Optical multichannelization is realized via time-division demultiplexing by the combination of tunable delay lines and dual-output Mach-Zehnder modulators. With the PDs converting the optical signal to electrical, electronic quantizers synchronized with the MLL quantize the signal to digital data. Deep learning data recovery is deployed in a computer with a central processing unit (CPU) and two graphical processing units (GPUs). It is worth noting that effective deep learning accelerators have been proposed via electronic (20–22) and optical (23, 24) schemes, inferring that deep learning data recovery could be carried out in real time in the near future. The computer works as a controller to the system, ensuring correct execution of data acquisition and processing in the training and testing procedures. Further detailed descriptions of the experimental setup can be found in Section 1 of Methods in Supplementary Materials.

We used deep neural networks, linearization nets, and matching nets to learn the pattern of system defects and perform data recovery from distorted data. Figure 2A depicts the basic model of these two kinds of neural networks. The first layer of the model, the input layer, accepts the original data and converts it to multiple feature channels by convolution. From the second layer, each layer is constructed with convolution manipulation and nonlinear activation (rectified linear units, ReLU, in this work). In addition, the two layers compose a residual block (25). At the end of each residual block, data are combined with those ahead of the residual block, finishing a residual short-cut. After several residual blocks, the output layer merges data from multiple feature channels and adds it to the original data. Figure 2B illustrates the convolutions of the neural networks, explaining why these purely convolutional neural networks are immune to data length variation and spectrum aliasing. The actuating range of convolutional windows are confined; therefore, the convolutional windows are trained to learn the local relations of the input and output. Hence, the trained convolutional windows are effective for sequences with arbitrary length. In addition, because the input and output sequences are given in the time domain, limitations of spectral analysis methods can be avoided in these neural networks.

**Validation of deep neural networks**

Based on the neural network model, we constructed the linearization nets and matching nets (neural network constructions are detailed in Section 2 of Methods in Supplementary Materials) and trained them with distorted data and their corresponding reference data (data acquisition, processing, and neural network training procedures are detailed in Section 3 of Methods in Supplementary Materials). In Fig. 3, the results of the training procedure and the effectiveness of nonlinearity corrections of different waveforms are presented. During training, some untrained sine data with different frequencies and amplitudes are used to test the inference validity of the neural networks (i.e., validation). Figure 3A depicts the variations of training loss and validation loss with the growth of training epochs. The loss here represents the absolute error of network output and reference data, meaning that the network output approaches the reference when loss becomes lower. Training loss is calculated on average among data in the training set and valid loss is calculated on average in the validation set. As can be seen, losses become lower when the training epochs grow and converge to a steady level. For better comprehension, the averaged signal-to-noise and distortion ratio (SINAD) is also calculated for the validation set. It converges to ∼47 dB, implying that linearization nets are viable in nonlinearity correction of untrained data spreading over the whole spectrum. As an example, an untrained signal in the time domain and frequency domain before and after the linearization nets is shown in Fig. 3B. The E/O-distorted waveform is corrected to a sine signal. In the frequency domain, we can clearly see that the harmonics due to the E/O nonlinearity have been eliminated. To check the broader applicability of linearization nets with other sine-like signals, we used dual-tone signals and linearly-frequency-modulated (LFM) signals to test the networks that were only trained by sine signals; some examples are given in Fig. 3C, D. Before linearization, dual-tone signals are distorted by E/O so that a series of distortions exist on the frequency spectrum; these distortions are effectively eliminated because of the trained linearization nets. The results demonstrate that the linearization nets can significantly extend the spurious-free dynamic range (SFDR) of the received signal amplitude, ensuring high accuracy of ADCs. Linearization nets also work effectively with LFM signals. In the spectra shown in Fig. 3D, second-order distortions are obviously suppressed. In the short-time Fourier transformation (STFT) spectrum, we obtained an approximate 26 dB improvement of the signal-to-distortion ratio before and after the neural networks. It is worth noting that the applied AWG has an accuracy of ∼6 effective numbers of bits (ENOB), meaning that noise and distortions in the LFM signal itself are relatively high, degrading the effectiveness of neural networks to some extent. More complete test results of linearization nets are presented in Fig. S2, S3 of Supplementary Materials, where the results show the reliability of deep neural networks in nonlinearity correction.

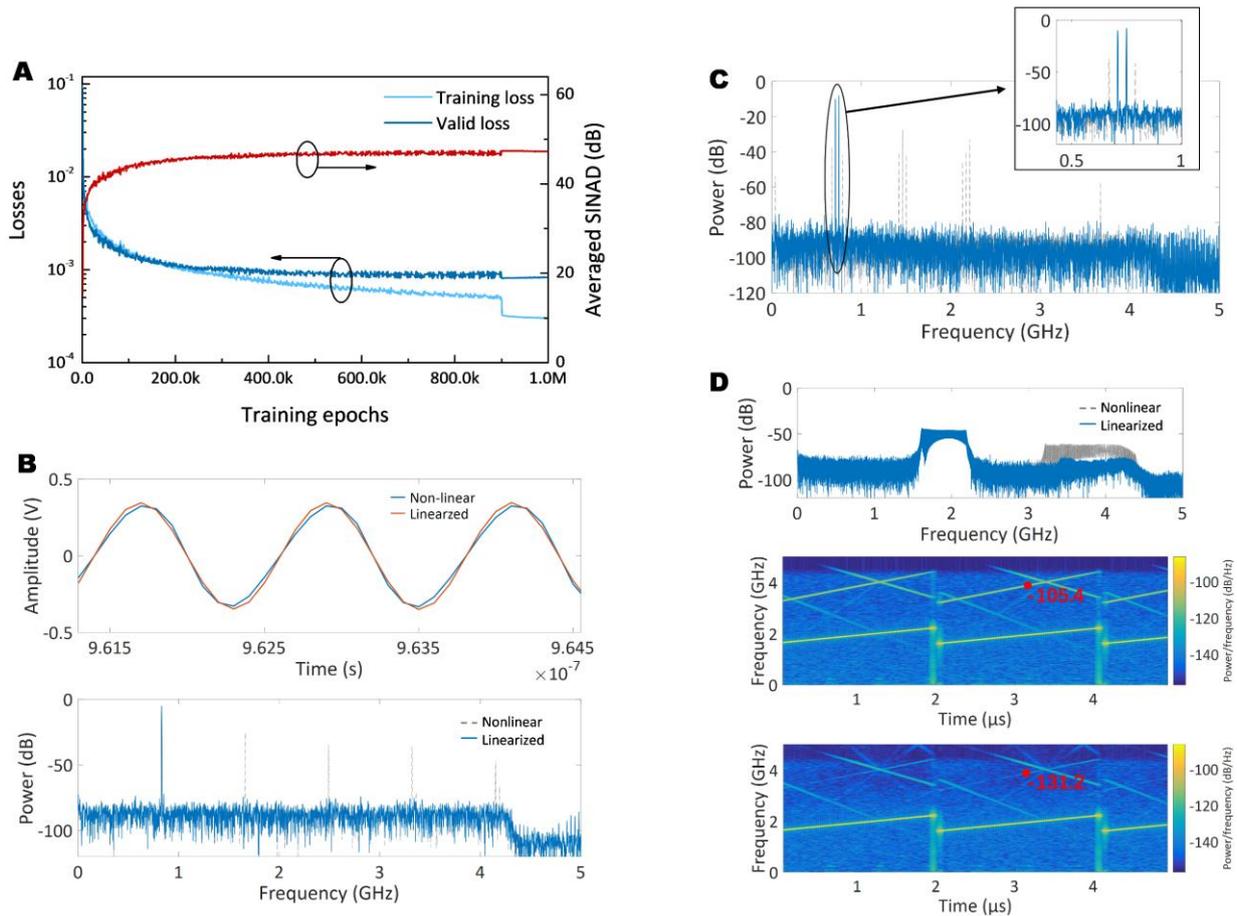

**Fig. 3. Results for validation of linearization nets.** (A) Performance of neural networks under training. The training loss and validation loss descend with the growth of training epochs. The training loss is obtained in the training set, by calculating the absolute error of network output and the reference data. Validation loss is the absolute error in the validation set that is not overlapped with the training set. For a better understanding of linearization performance, we use the red curve to show the averaged SINAD in the validation set. (B) An example of nonlinear distortion elimination. The test sine signal is 833 MHz, ± 0.36 V. The upper subplot shows the time domain and lower subplot is the frequency domain. (C) and (D) Examples of neural networks applicability with untrained dual-tone signals and LFM signals. In (C), similarly, the gray dashed curve and the blue curve represent the data before and after linearization nets, respectively. The dual-tone signals are of frequencies 712 MHz and 752 MHz. In (D), frequency spectrum and STFT spectra are presented for illustrating distortion elimination of an LFM signal. We mark the values at the same location in the upper and lower STFT spectra. The LFM signal is in the 1.60–2.20 GHz range.

The results demonstrated in Fig. 4 show the effectiveness of matching nets. We consider each reference data of the linearization nets as a single input of matching nets and train the network with reference interleaved data (data acquisition, processing, and neural network training procedures are detailed in Section 3 of Methods in Supplementary Materials). Figure 4A shows the results of training of matching nets. With the epochs growing, training and validation losses reduce and converge to a steady level, and the averaged SINAD also approaches the noise-floor-limited level, ~46 dB. An example of the sine signal is given in Fig. 4B. In the time-domain plot, channel mismatches leave errors on the interleaved data, and bring in the mismatch distortions on the frequency spectrum. With the trained matching nets, the errors are corrected and the mismatch distortions are compensated effectively. Furthermore, the matching nets can accomplish channel mismatch compensation of broadband signals. Figure 4C depicts an example of the compensation of a mismatch-distorted LFM signal. On the right side of the spectrum resides the broadband distortion introduced by the channel mismatch. The matching nets eliminate it with high quality, as can be explicitly seen in the following STFT spectra. Since the number of channels determines the sampling rate multiplication and electronic burden releasing, the matching nets should also be compatible with multichannel data interleaving. To ensure the expandability of the constructed matching nets, simulations were conducted with a varied number of channels (detailed in Section 3 of Methods in Supplementary Materials). Corresponding to each number of channels, ten mismatch degrees were randomly selected referring to the mismatch degree in the two-channel experiment. For every mismatch degree for every number of channels, we trained the matching nets to interleave mismatched data; the results are shown in Fig. 4D. The averaged SINAD in the validation set converge at around 46 dB, implying that matching nets are adaptive with different numbers of channels and different mismatch degrees. The above-mentioned results, together with additional test results (Fig. S4, S5 in Supplementary Materials), provide validation of the matching nets in channel mismatch compensation.

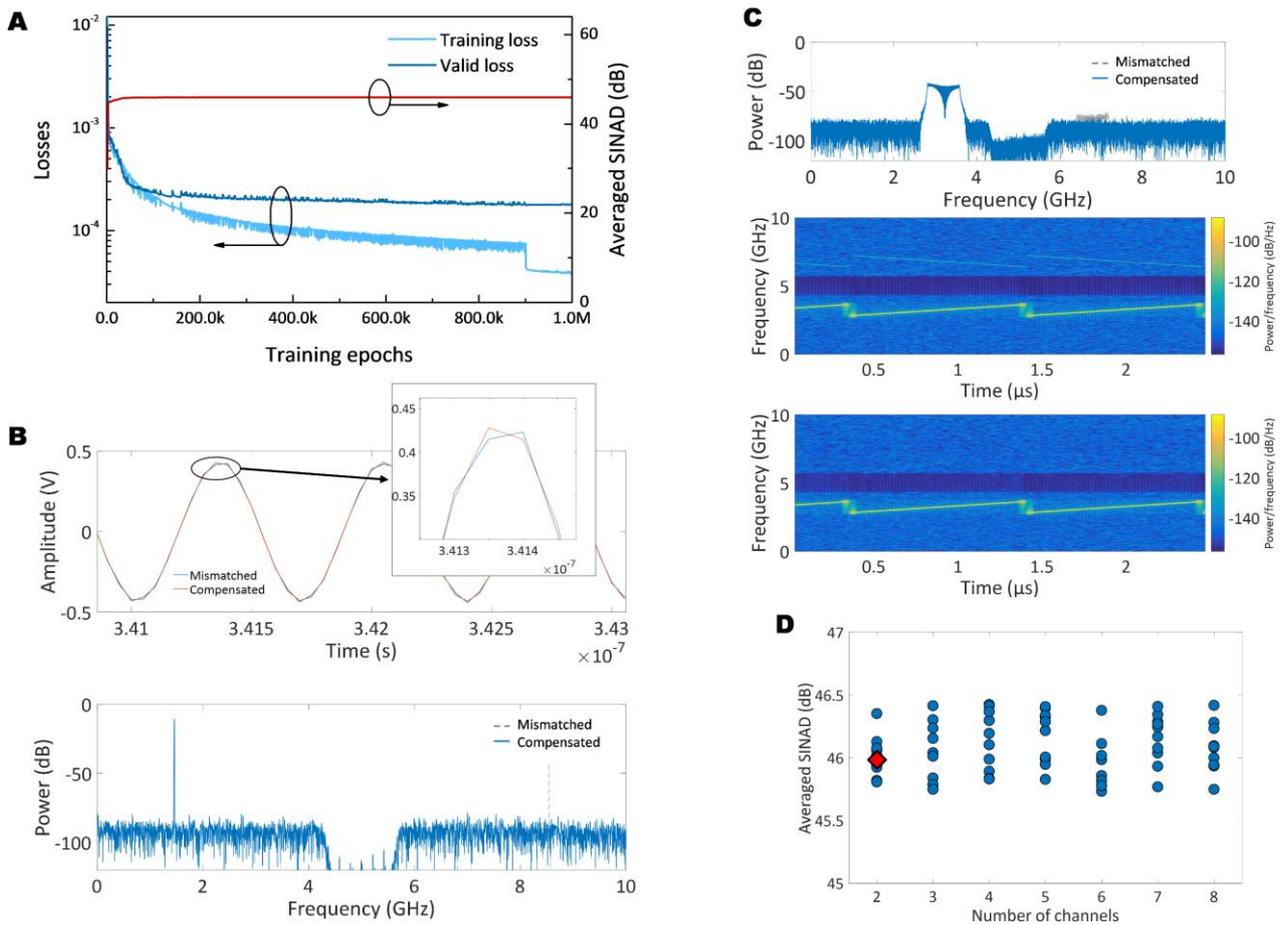

**Fig. 4. Results for the validation of matching nets.** (A) The losses are descending with the growth of training epochs. The red curve represents the averaged SINAD in the validation set. (B) An example of channel mismatch compensation with matching nets. The test signal is 1.468 GHz and ± 0.42 V. The time domain and frequency domain plots depict the effectiveness of matching nets. We zoom in the time domain plot for a better view of the error between mismatched and compensated data. (C) Frequency spectra and STFT spectra are shown to illustrate the neural network's applicability with unstrained LFM signals. The signal is in the 1.40-1.80 GHz range. (D) Simulation results of the expandability of matching nets for greater number of channels. We tried different numbers of channels (2 to 8) and different mismatch degrees (referring to the mismatch degree in our experiment). For every combination of number of channels and mismatch degree, we conducted 1 million training epochs and marked the averaged SINAD in the figure. The red diamond denotes the experimental result.

**Performance characterization**

As the effectiveness of the neural networks was completely demonstrated with different signal formats

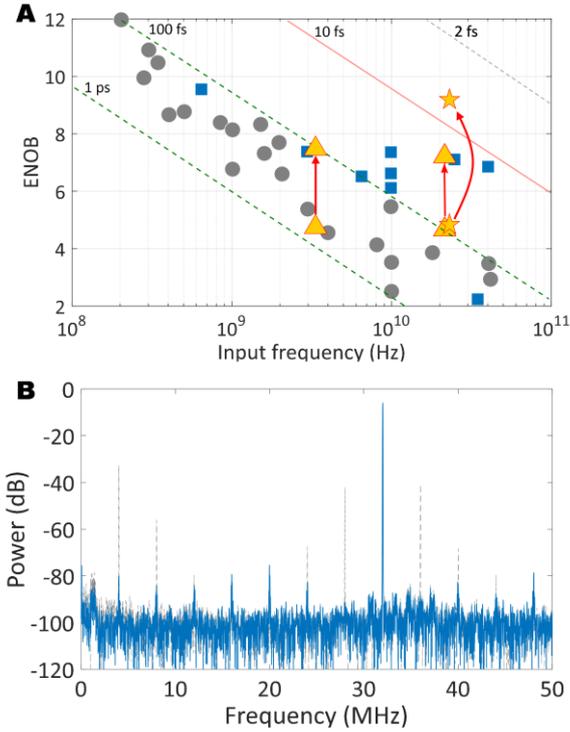

**Fig. 5. Performance characterization of the proposed analog-to-digital conversion architecture.** (A) Walden plot filling for state-of-the-art ADC systems (data sampled from (3)). Gray dots show the performances of electronic ADCs and blue rectangles show photonics-assisted ADC performances. Yellow triangles illustrate the performance improvement of the experimental setup for ADC before and after the application of deep learning (the signal frequencies are 3.44 GHz and 21.13 GHz, respectively). The yellow stars mark the ENOB results for the further experiment using a 100-MHz PMLL and 100 MS/s data acquisition board (the signal frequency is 23.332 GHz). (B) Frequency spectrum before and after nonlinearity cancellation by the linearization nets using the PMLL and data acquisition board in the experimental setup. The input frequency is 23.332 GHz.

(Fig. 5), we further characterize the performance enhancement of the experimental 20-GS/s ADC setup and compare it with state-of-the-art commercial and in-lab ADCs by using the Walden plot. We test the Nyquist sampling of a 3.44-GHz sine signal and the subsampling of a 21.13-GHz sine signal with the experimental setup. Before the test signal is sampled and quantized, the training procedure is performed with the above-mentioned training set. In principle, a signal of frequency 21.13 GHz will be subsampled to 1.13 GHz so that the trained neural networks can be adaptive when directly sampling high-frequency signals. In Fig. 5A, two results are marked for each test signal; before data recovery by the deep neural networks, the SINAD is deteriorated severely because of the E/O nonlinearity and channel mismatch distortions, resulting in 4.66 ENOB with an input frequency of 3.44 GHz, and 4.53 ENOB with 21.13 GHz. After two cascaded steps of data recovery, the SINAD increases significantly, reaching 7.28 ENOB with an input frequency of 3.44 GHz, and 7.07 ENOB with 21.13 GHz. The accuracy performance does not surpass that of the state-of-the-art ADC because it is realized with inferior electronic quantization (the oscilloscope), whose quantization noise heavily limits further accuracy enhancement. To realize the ultimate accuracy of the neural networks, we conducted a further experiment with a 100-MHz mode-locked laser with 2 femtosecond timing jitter and a 100-MHz high-accuracy data acquisition board (detailed in Section 5 of Methods in Supplementary Materials). Although the sampling rate is low, this experimental setup provides an ultralow noise level, demonstrating the accuracy capability of the neural networks. We tested the accuracy capability of linearization nets, and the capability of matching nets does not differ much. The ENOB results are also shown in Fig. 5A. With the elimination of nonlinear distortions, the ENOB had been enhanced from 4.57 to 9.24 with an input frequency of 23.332 GHz. Figure 5B shows the spectrum of the linearized 23.332-GHz signal, demonstrating that nonlinear distortions are effectively eliminated and the SFDR is markedly enlarged. By testing the signals over the whole frequency range, the SFDR is characterized above 68 dB and is 71 dB on average (ENOB and SFDR characterizations are described in Supplementary Materials).

**Conclusions**

In this work, a revolutionary analog-to-digital conversion architecture including three cascaded parts was proposed for the next generation of broadband high-speed high-accuracy ADC design. A photonic front-end provides the feasibility of directly sampling broadband RF as well as multiplication of the sampling rate. Although the system defects of the photonic front-end pervade the quantized data, degrading the accuracy, deep learning methods provide a succinct way to learn the patterns of system defects and recover the distorted data. Together with electronic high-accuracy quantization, this architecture has exploited the advantages of every component to overcome the present ADC bottleneck. To demonstrate the reliability of the proposed architecture, we set up a 20-GS/s ADC with two time-divided demultiplexed channels. With the demonstration of the effectiveness of deep neural networks, the holistic ADC offers high accuracy in Nyquist sampling and subsampling, presenting a remarkable performance compared with the state-of-art commercial and in-lab ADCs. In further simulations and experiments, we demonstrated the expandability for greater number of multiplexed channels and the accessibility to higher accuracy and higher dynamic range. Therefore, the proposed deep-learning-powered analog-to-digital conversion architecture is believed to be reliable with regard to future ADC design requirements if less noisy electronic quantization and greater number of channels (demonstrated in Fig. 4D) are adopted, offering high-speed, broadband, and high-accuracy opportunities for next-generation information systems to greatly enhance their capability.

**References and Notes:**

1. J. G. Andrews, S. Buzzi, W. Choi, S. V. Hanly, A. Lozano, A. C. K. Soong, J. C. Zhang, What will 5G be? IEEE J. Sel. Area Comm. 32, 1065-1082 (2014).
2. W. Zou, H. Zhang, X. Long, S. Zhang, Y. Cui, J. Chen, All-optical central-frequency-programmable and bandwidth-tailorable radar. Sci. Rep. 6, 19786 (2016).
3. P. Ghelfi, F. Laghezza, F. Scotti, G. Serafino, A. Capria, S. Pinna, D. Onori, C. Porzi, M. Scaffardi, A. Malacarne, V. Vercesi, E. Lazzeri, F. Berizzi, A. Bogoni, A fully photonics-based coherent radar system. Nature 507, 341-345 (2014).
4. G. C. Valley, Photonic analog-to-digital converters. Opt. Express 5, 1955-1982 (2007).
5. A. Khilo, S. J. Spector, M. E. Grein, A. H. Nejadmalayeri, C. W. Holzwarth, M. Y. Sander, M. S. Dahlem, M. Y. Peng, M. W. Geis, N. A. Dilello, J. U. Yoon, A. Motamedi, J. S. Orcutt, J. P. Wang, C. M. Sorace-Agaskar, M. A. Popovic, J. Sun, G. Zhou, H. Byun, J. Chen, J. L. Hoyt, H. I. Smith, R. J. Ram, M. Perrott, T. M. Lyszczarz, E. P. Ippen, F. X. Kartner, Photonic ADC: overcoming the bottleneck of electronic jitter. Opt. Express 20, 4454-4469 (2012).
6. J. Yao, Microwave photonics. J. Lightwave Technol. 27, 314-335 (2009).
7. P. W. Juodawlkis, J. C. Twichell, G. E. Betts, J. J. Hargreaves, R. D. Younger, J. L. Wasserman, F. J. O'Donnell, K. G. Ray, R. C. Williamson, Optically sampled analog-to-digital converters. IEEE T. Microw. Theory. 49, 1840-1853 (2001).
8. G. Yang, W. Zou, X. Li, J. Chen, Theoretical and experimental analysis of channel mismatch in time-wavelength interleaved optical clock based on mode-locked laser. Opt. Express 23, 2174-2186 (2015).
9. G. Yang, W. Zou, L. Yu, K. Wu, J. Chen, Compensation of multi-channel mismatches in high-speed high-resolution photonic analog-to-digital converter. Opt. Express 24, 24061-24074 (2016).
10. Y. LeCun, Y. Bengio, G. Hinton, Deep learning. Nature 521, 436-444 (2015).
11. A. Krizhevsky, I. Sutskever, G. E. Hinton, Image classification with deep convolutional neural networks. Adv. Neural Inf. Process. Syst. 25, 1097-1105 (2012).
12. J. Tompson, A. Jain, Y. LeCun, C. Bregler, Joint training of a convolutional network and a graphical model for human pose estimation. Adv. Neural Inf. Process. Syst. 27, 1799-1807 (2014).
13. M. Anthimopoulos, S. Christodoulidis, L. Ebner, A. Christe, S. Mougiakakou, Lung pattern classification for interstitial lung disease using a deep convolutional neural network. IEEE T. Med. Imaging 35, 1207-1216 (2016).
14. D. Silver, J. Schrittwieser, K. Simonyan, I. Antonoglou, A. Huang, A. Guez, T. Hubert, L. Baker, M. Lai, A. Bolton, Y. Chen, T. Lillicrap, F. Hui, L. Sifre, G. V. D. Driessche, T. Graepel, D. Hassabis, Mastering the game of Go without human knowledge. Nature 550, 354-359 (2017).
15. X. Lu, Y. Tsao, S. Matsuda, C. Hori, Speech enhancement based on deep denoising autoencoder. In Interspeech 2013, 436-440 (2013).
16. J. Xie, L. Xu, E. Chen, Image denoising and inpainting with deep neural networks. Adv. Neural Inf. Process. Syst. 25, 341-349 (2012).
17. Y. Riverson, Z. Gorocs, H. Gunaydin, Y. Zhang, H. Wang, A. Ozcan, Deep learning microscopy. Optica 4, 1437-1443.
18. B. Zhu, J. Z. Liu, S. F. Cauley, B. R. Rosen, M. S. Rosen, Image reconstruction by domain-transform manifold learning. Nature 555, 487-492 (2018).
19. L. Pierno, A. M. Fiorello, A. Bogoni, P. Ghelfi, F. Laghezza, F. Scotti, S. Pinna, Optical switching matrix as time domain demultiplexer in photonic ADC. In Proc. the 8th Euro. Microw. Integrated Circuits Conf. 41-44 (2013).
20. A. Coates, B. Huval, T. Wang, D. J. Wu, A. Y. Ng, Deep learning with COTS HPC systems. International Conf. Machine Learning, 1337-1345 (2013).
21. N. P. Jouppi, C. Young, N. Patil, D. Patterson, G. Agrawal, R. Bajwa, S. Bates, S. Bhatia, N. Boden, A. Borchers, R. Boyle, P. Cantin, C. Chao, C. Clark, J. Coriell, M. Daley, M. Dau, J. Dean, B. Gelb, T. V. Ghaemmaghami, R. Gottipati, W. Gulland, R. Hagmann, C. Richard Ho, D. Hogberg, J. Hu, R. Hundt, D. Hurt, J. Ibarz, A. Jaffey, A. Jaworski, A. Kaplan, H. Khaitan, D. Killebrew, A. Koch, N. Kumar, S. Lacy, J. Laudon, J. Law, D. Le, C. Leary, Z. Liu, K. Lucke, A. Lundin, G. MacKean, A. Maggiore, M. Mahony, K. Miller, R. Nagarajan, R. Narayanaswami, R. Ni, K. Nix, T. Norrie, M. Omernick, N. Penukonda, A. Phelps, J. Ross, M. Ross, A. Salek, E. Samadiani, C. Severn, G. Sizikov, M. Snelham, J. Souter, D. Steinberg, A. Swing, M. Tan, G. Thorson, B. Tian, H. Toma, E. Tuttle, V. Vasudevan, R. Walter, W. Wang, E. Wilcox, D. H. Yoon, In-datacenter performance analysis of tensor processing unit. In Proc. 2017 International Symposium on Computer Architecture 1-12 (2017).
22. S. Ambrogio, P. Narayanan, H. Tsai, R. M. Shelby, I. Boybat, C. Nolfo, S. Sidler, M. Giordano, M. Bodini, N. C. P. Farinha, B. Killeen, C. Cheng, Y. Jaoudi, G. W. Burr. Equivalent-accuracy accelerated neural network training using analog memory. Nature 558, 60-67 (2018).
23. Y. Shen, N. C. Harris, S. Skirlo, M. Prabhu, T. Baehr-Jones, M. Hochberg, X. Sun, S. Zhao, H. Larochelle, D. Englund, M. Soljacic, Deep learning with coherent nanophotonic circuits. Nat. Photon. 11, 441-446 (2017).
24. X. Lin, Y. Rivenson, N. T. Yardimci, M. Veli, Y. Luo, M. Jarrahi, A. Ozcan, All-optical machine learning using diffractive deep neural networks. Science 361, 1004-1008 (2018).
25. K. He, X. Zhang, S. Ren, J. Sun, Deep residual learning for image recognition. In Proc. IEEE Conf. Computer Vision and Pattern Recognition 770-778 (2016).
26. S. C. Park, M. K. Park, M. G. Kang, Super-resolution image reconstruction: a technical overview. IEEE Signal Process. Mag. 20, 21-36 (2003).

27. W. Shi, J. Caballero, F. Huszar, J. Totz, A. P. Aitken, R. Bishop, D. Ruechert, Z. Wang, Real-time single image and video super-resolution using an efficient sub-pixel convolutional neural network. IEEE Conf. Computer Vision Pattern Recognition, 1874-1883 (2016).
28. D. Han, J. Kim, J. Kin, Deep pyramidal residual networks. http://arxiv.org/abs/1610.02915 (2016).
29. K. He, X. Zhang, S. Ren, J. Sun, Identity mapping in deep residual networks. In Proc. European Conf. Computer Vision (2016), Springer, Cham, 630-645.
30. K. Zhang, W. Zuo, Y. Chen, D. Meng, L. Zhang, Beyond a Gaussian denoiser: residual learning of deep CNN for image denoising. IEEE T. Image Process. 26, 3142-3155 (2017).
31. V. N. V. S. Prakash, K. S. Peasad, T. J. Prasad, Deep learning approach for image denoising and image demosaicing. International J. Computer Applications 168, 18-26 (2017).
32. M. Long, Y. Cao, J. Wang, M. Jordan, Learning transferable features with deep adaptation networks, 32nd International Conf. Machine Learning, 97-105 (2015).
33. R. Collobert, J. Weston, A unified architecture for natural language processing: deep neural networks with multitask learning, 25th International Conf. Machine Learning, 160-167 (2008).
34. S. Klein, J. P. W. Pluim, M. Staring, Adaptive stochastic gradient descent optimization for image registration. International J. Computer Vision 81, 227-239 (2009).

**Funding:** National Natural Science Foundation of China (grant no. 61822508, 61571292, 61535006); **Author contributions:** S. X. and W. Z. conceived the research; S. X., X. Z., B. M., J. C., and L. Y. contributed to the experiments; S. X. processed the data; S. X. and W. Z. prepared the manuscript; W. Z. initiated and supervised the research. **Competing interests:** W. Z, S. X., and J. C. are the inventors of a patent application on the ADC architecture. **Data and materials availability:** All data are available in the main text or the supplementary materials.

**Supplementary Materials:**
Materials and Methods
Supplementary text
Figures S1-S5
References (26-34)

Supplementary Materials for

# Analog-to-digital Conversion Revolutionized by Deep Learning

Shaofu Xu[1], Xiuting Zou[1], Bowen Ma[1], Jianping Chen[1], Lei Yu[1], Weiwen Zou[1, *]

[1]State Key Laboratory of Advanced Optical Communication Systems and Networks, Department of Electronic Engineering, Shanghai Jiao Tong University, Shanghai 200240, China.

*Correspondence to: wzou@sjtu.edu.cn.

## Methods

1. <u>Experimental setup of the 20-GS/s ADC</u>

Based on the proposed analog-to-digital conversion architecture, we set up a 2-channel 20 GS/s ADC for the validation (the experimental setup is shown in Fig. S1). We implemented the photonic front-end with an actively mode-locked laser (AMLL, CALMAR PSL-10-TT), a microwave generator (MG1, KEYSIGHT E8257D), a Mach-Zehnder modulator (MZM, PHOTLINE MXAN-LN-40), and a two-channel time-division demultiplexer. Driven by the MG1 at a frequency of 20 GHz, the AMLL emitted optical pulses at 20 GHz repetition rate. As a reference, the measured timing jitter of the AMLL output optical pulse was around 26.5 fs. The MZM adopted had a bandwidth of 40 GHz, guaranteeing the reception of high-frequency broadband signals. In the MZM, the optical pulse train from the AMLL was amplitude modulated by the signal to be sampled, so the signal was sampled with a fixed interval. The two-channel time-divided demultiplexer consisted of a tunable delay line (TDL, General Photonics MDL-002) whose tuning accuracy was 1 ps, a dual-output Mach-Zehnder modulator (DOMZM, PHOTLINE AX-1x2-0MsSS-20-SFU-LV) of low quadrature voltage $V_\pi = 3.5V$, and two identical custom-built PDs of 10 GHz bandwidth. For demultiplexing the optical pulse train into two channels, the custom-built frequency divider transferred the 20 GHz signal from the MG1 to 10 GHz and drove the DOMZM. The DOMZM was biased at its quadrature point and the driving 10 GHz signal was adjusted to match full $V_\pi$ of the DOMZM. Subsequently, we adjusted the TDL, letting one optical pulse of two adjacent pulses pass through the DOMZM at its maximal transmission rate, and allowing another pulse to pass through the MZM at its minimal transmission rate. Therefore, the optical pulse train was demultiplexed into 2 channels. To evaluate the effectiveness of the demultiplexer, we used a 50-GHz PD (u2t XPDV2150R) and a sampling oscilloscope (KEYSIGHT DCA-X 86100D) to test the demultiplexed optical pulses. During the electronic quantization, a multi-channel real-time oscilloscope (OSC, KEYSIGHT DSO-S 804A) was adopted as the quantizer; it had 10 GS/s sampling speed and 4 channels. As a reference, we measured the ENOB of the OSC at 7.4 maximally. The OSC was synchronized by the MG1 to keep the quantization clock synchronized with the AMLL. In the following deep learning data recovery, a computer with a CPU core (Intel CORE i7-7700K) and two GPUs (NVidia GTX 1080ti) was programmed to construct linearization nets and matching nets. We used Tensorflow (v1.6) in Python as the framework to program the neural networks and LabVIEW to program the interfaces between the computer and instruments. To generate the training signals, another microwave generator (MG2, KEYSIGHT N5183B) was adopted. Controlled by the computer, it generated the signals to be sampled and input them into MZM. Since the output signal of MG2 contained harmonics other than standard sine, a series of custom-built low pass filters (LPFs) were prepared to cancel the harmonics, ensuring that the output signal of

MG2 was clean. For the validation of the ADC in untrained sine-alike signal applicability, we applied dual-tone signals and LFM signals as input to the ADC. The dual-tone signals were generated by the combination of MG2 and another microwave generator (MG3, Rhode & Schwarz SMA 100A), and the LFM signals were generated via an arbitrary waveform generator (AWG, KEYSIGHT M9502A).

2. Implementations of deep neural networks

Inspired by image de-noising, inpainting, and super-resolution (26, 27), the tasks of nonlinearity cancellation and mismatch compensation only need the neural networks to manipulate local data rather than memorizing the whole data sequence. Therefore, we could construct the neural networks to be purely convolutional, which bought significant advantages for the ADC application (i.e., immunity to data length variation and frequency in spectrum aliasing). The neural networks were composed by a residual learning scheme (25) and the linearization nets were comprised with an input layer, four residual blocks, and an output layer. The input layer was a convolution layer convert of one input channel to 32 feature channels, represented by:

$$Y_j = X_i * W_{ij} + b_j, j = 1, 2, ..., 32$$

The input channel $X_i (i=1)$ consisted of an input data sequence convoluted with the $j$-th convolution window $W_{ij}$, whose window width was 3, in the 'SAME' manner (padding the head and the tail of the input sequence with zeros; hence, output is of the same length as input). We then added the $j$-th bias $b_j$ to get the j-th feature channel $Y_j$. In the following residual blocks, two convolution and activation layers were included. Each layer of convolution and activation was represented by:

$$Y_j = \text{ReLU}(\sum_{i=1}^{N} X_i * W_{ij} + b_j), j = 1, 2, ..., J$$

Different from the input layer, this layer has a 'ReLU' manipulation, which means $\text{ReLU}(x) = \max\{0, x\}$. We changed the number of output feature channels $J$ as the pyramid structure (28). At the end of each residual block, $J$ = 34, 38, 44, or 52, respectively. As the output data of each residual block should be added to the input of the residual block, but were unmatched on feature channel numbers, we used an additional convolution layer (the window width was 1) to convert the channel number of the input to match the channel number of the output (29). The output layer was similar to the input layer of the calculation formula, but it converted the 52 feature channels to one output data sequence. By adding the output data sequence with the original input data sequence (30, 31), the output of the linearization nets was obtained. As for the matching nets, the original input data were several sequences from different quantization channels. So, in the input layer of matching nets, we conducted interleaving after individual convolutions as follows:

$$Y_j^m = X_i^m * W_{ij}^m + b_j^m, j = 1, 2, ..., 32, m = 1, 2$$
$$Y_j = \text{ITL}(Y_j^1, Y_j^2)$$

The 'ITL' manipulation is interleaving, constructing the result sequence $Y_j$ by alternately picking the data in $Y_j^1$ and $Y_j^2$ (i.e., $Y_j[1], Y_j[2], Y_j[3], Y_j[4], Y_j[5]... = Y_j^1[1], Y_j^2[1], Y_j^1[2], Y_j^2[2], Y_j^1[3]...$). For each input data sequence, we calculated 32 feature channels and then used interleaving to construct 32 interleaved feature channels. The interleaved feature channels were double the length of the input data sequence. The following part of the 'Matching nets' was the same as that of the 'Linearization nets', with four residual blocks and an output layer.

3. Data acquisition, processing, and neural network training

In the experimental demonstration of the effectiveness of the proposed analog-to-digital conversion architecture, 417 sine signals with varied frequencies and amplitudes, dual-tone signals with different frequencies, and LFM signals with different frequencies and bandwidths were sampled by the experimental setup in order to construct the training dataset and the validation dataset. The deep neural networks in this work were trained with sine inputs; acceptable signal waveforms were sine-like. In future work, by using new datasets and new training methods (32, 33), the neural networks will be applicable in more complicated waveforms. Since the sampling rate of the experimental setup was 20 GS/s, the frequencies of the sampled sine signals were randomly selected but uniformly distributed within the Nyquist bandwidth of 0~10 GHz. As the adopted real-time oscilloscope has a built-in bandwidth limit of 4.2 GHz, we discard the frequencies from 4 GHz to 6 GHz. By linking appropriate LPFs on the output of MG2, second order or high order harmonics of the output signals were eliminated. A LabVIEW program was developed to control MG2 in order to emit amplitude/frequency-varying signals. The amplitudes were also randomly selected and uniformly distributed within −2 to 15 dBm. The dual-tone signals were generated by the combination of MG2 and MG3, and the LFM signals were generated by AWG. Appropriate filters were also used in dual-tone and LFM signals avoiding harmonics residing in the generated signals. The data processing yielded the training set and validation set by obtaining original/reference data pairs. To train the linearization nets, we took the distorted results as the original data and calculated the reference data for every distorted result. By removing the nonlinear harmonics by frequency domain analysis and adding the harmonics power to the signal power, the processed signal was regarded as the reference data. The LFM signals whose spectra were not aliased were processed as such since frequency domain analysis is inappropriate for aliased spectra. This data processing was performed using MATLAB codes. To train the matching nets, the original data were the reference data for 'Linearization nets' gained from the abovementioned processing, and the reference data were the recovered interleaved data. Frequency domain manipulation was also used for the reference data processing, removing channel mismatch distortions and adding the power to signal power. By dividing 367 data pairs as the training set and 50 data pairs as the validation set, we conducted neural network training by minimizing the loss in the training set:

$$Loss(\Theta) = \frac{1}{L}\sum_{l=1}^{L}|Y_l^{\Theta} - Y_l^{REF}|$$

We reconfigured the parameters of the neural networks $\Theta$ by adopting minimization algorithms to minimize the averaged absolute difference between the output of the neural networks $Y^{\Theta}$ and the reference data $Y^{REF}$. The minimization algorithm used in this work was adaptive gradient descendance (34) with backpropagation. Here, $L$ represents the length of data sequences, 1000 in the linearization nets and 2000 in the matching nets. In total, 1 million training epochs were conducted for every neural network and we calculated the loss in the validation set every 1000 epochs.

4. Simulation of 'matching nets' applicability in multi-channels

Using the experimental setup, the validity of matching nets was demonstrated using 2-channel data interleaving. For further sampling rate multiplication, we used the simulation results to show the capability of matching nets in multi-channel data interleaving. The simulation was conducted in the following steps:

(1) Take the reference data of the matching nets (calculated as per Section 3) as the reference data in the simulation. The original data will be calculated from the reference data by adding some mismatches.
(2) Divide the reference data to N channels (N varies from 2 to 8). This procedure is inverse to

interleaving, and allocates data into different channels alternately.
(3) Add some channel mismatches to the data in each channel. The mismatch degree in the experimental setup is about 7 ps; therefore, mismatch degrees in the simulations are randomly selected around 7 ps. This data processing can be done using MATLAB codes.
(4) Use the artificial mismatched channels and reference data to train the matching nets for 1 million epochs and record the converged values.
(5) Change the mismatch degrees and number of channels N and repeat step (2) ~ (4).
For each number of channels, 10 mismatch degrees are tried and recorded in Fig. 4d.

5. Supplementary experiment using low-jitter MLL and high accuracy data acquisition board
To manifest the high accuracy of the neural networks and demonstrate the potential of the proposed analog-to-digital conversion architecture in future high-dynamic high-accuracy applications, an ultralow-jitter MLL (Menlo Systems LAC-1550) was adopted to replace the AMLL and a high-accuracy electronic data acquisition board (Texas Instrument ADC16DX37EVM) replaced the OSC. The nominal timing jitter of the PMLL was less than 2 fs and the ENOB of the data acquisition board was 9.37, facilitating ultralow noise floor. The repetition rate of the PMLL was 100 MHz and the sampling rate of the data acquisition board was 100 MHz. Since the Nyquist bandwidth of 100 MS/s ADC is 50 MHz, to acquire the training set and validation set, we controlled the MG1 to generate signals from 400 MHz to 450 MHz to match the passband of the low pass filter, which could suppress the harmonics of signals from 330 MHz to 500 MHz. The PD was replaced with a 300-MHz PD to avoid extra thermal noise. In total, 274 sine data were obtained, wherein 244 were selected as the training set and 30 of them were the validation set. The data acquisition, processing, and neural network training methods were similar to those detailed in Section 3. After training, this setup was used to conduct subsampling of the 23.333-GHz signal.

## Supplementary Text

1. Nyquist sampling and subsampling

Following the Nyquist sampling law, the bandwidth of the sampled signal is limited by half of the sampling rate:

$$Bandwidth \leq \frac{SamplingRate}{2}$$

If the frequency components of a signal obey the bandwidth limit, the signal can be sampled and quantized with its original information maintained; this is so-called Nyquist sampling. If a signal has a high-frequency carrier but its bandwidth follows the Nyquist sampling law, it can be aliased and quantized to its baseband. The information is still maintained; this is called subsampling. Therefore, any signal that does not exceed the bandwidth limitation can be sampled to baseband without information loss.

2. ENOB and SFDR characterizations

We conducted performance characterizations of our experimental setup with the IEEE standards. For an ADC system, single tone (sine) signals are used for ENOB and SFDR characterizations.

When the signals to be sampled are of single tone, ENOB can be represented by the ratio of the power of the signal to the power of all the noise and distortions:

$$ENOB = \frac{1}{6.02} \cdot \left( 10\log_{10}\left( \frac{P_{signal}}{P_{noise} + P_{distortions}} \right) - 1.76 \right) = \frac{SINAD - 1.76}{6.02}$$

The SINAD here was calculated in dB using the MATLAB "sinad()" function.

The SFDR of an ADC is defined as the ratio of the power of the signal to the power of the largest harmonic or distortion:

$$SFDR(\text{dB}) = 10\log_{10}\left(\frac{P_{signal}}{P_{max\_harm} \text{ or } P_{max\_distortion}}\right)$$

The power of signals and harmonics or distortions are calculated from the spectra after adding a Blackman window.

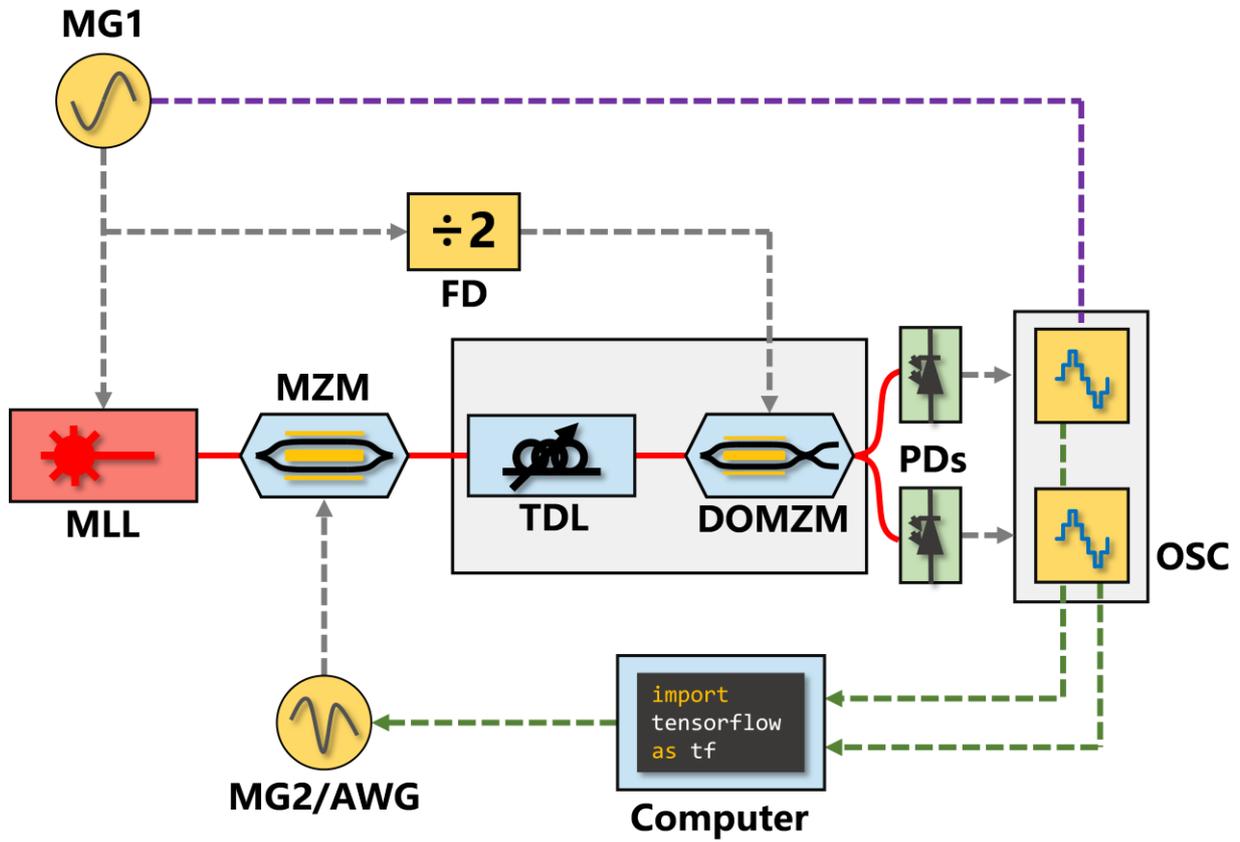

**Fig. S1. Experimental setup of the 20-GS/s ADC.** MG, microwave generator; FD, frequency divider; MLL, mode-locked laser; MZM, Mach–Zehnder modulator; TDL, tunable delay line; DOMZM, dual-output Mach–Zehnder modulator; PD, photon detector; OSC, oscilloscope; AWG, arbitrary waveform generator.

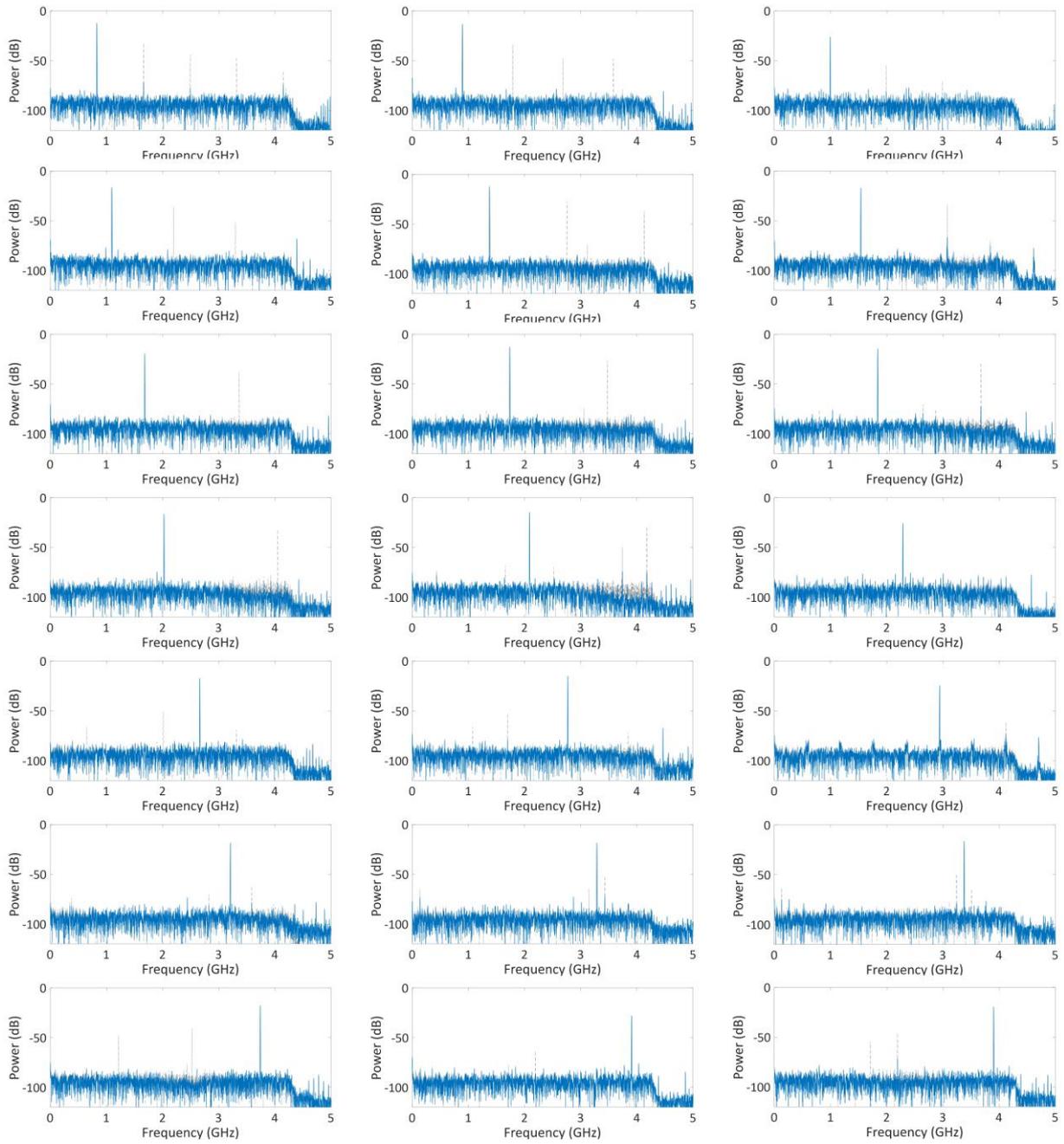

**Fig. S2. Sine signals results before and after linearization nets.** These data are randomly chosen in the validation set. Grey dashed curve is the original sampled and quantized data. Blue solid curve is the linearization nets recovered data.

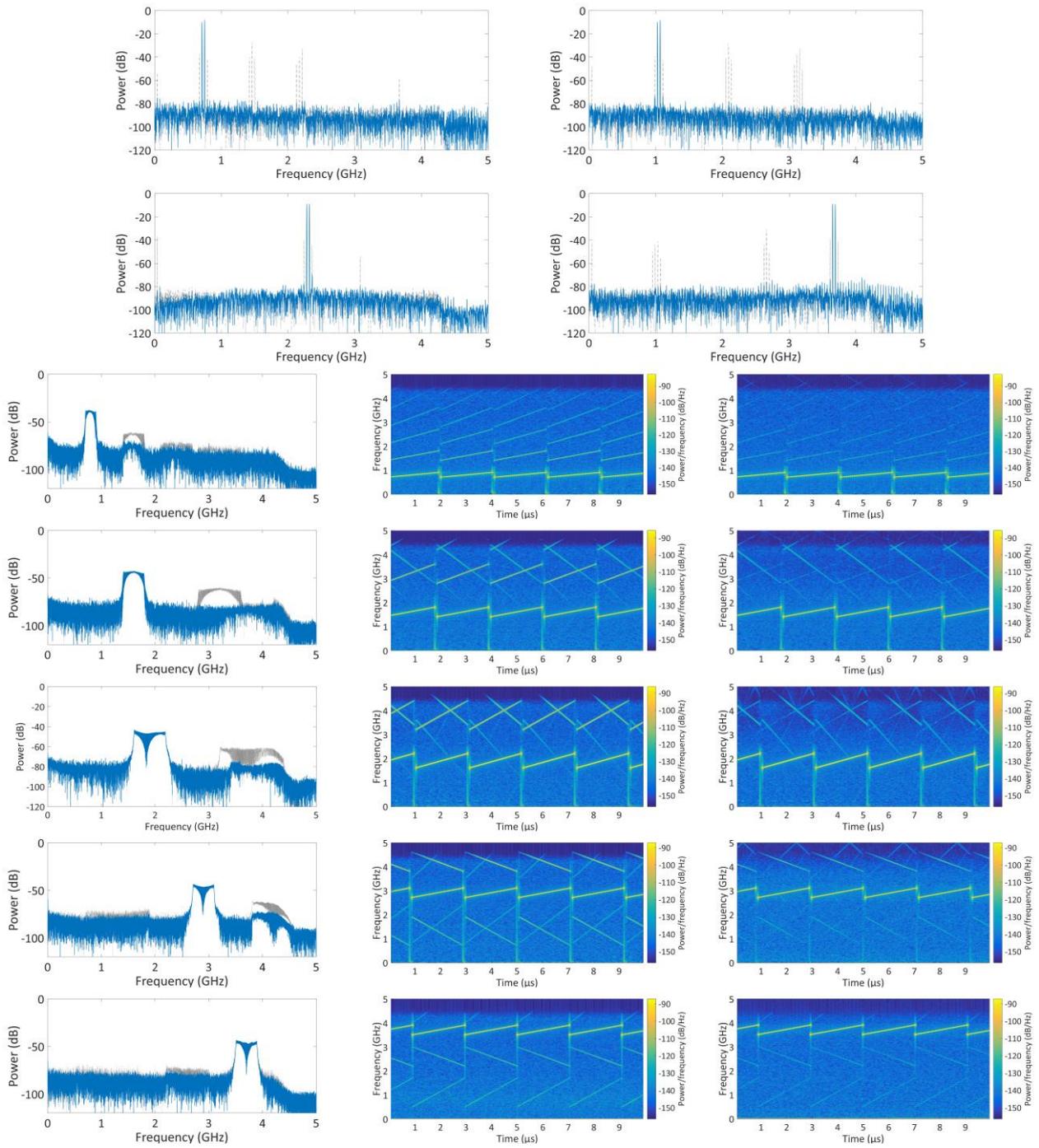

**Fig. S3. Dual-tone and LFM signals results before and after linearization nets.** In all frequency spectra, grey dashed curves represent data before linearization and blue solid curves denote linearization nets recovered data. STFT plots are also given for LFM signals: the left and right columns are data before and after the linearization nets, respectively.

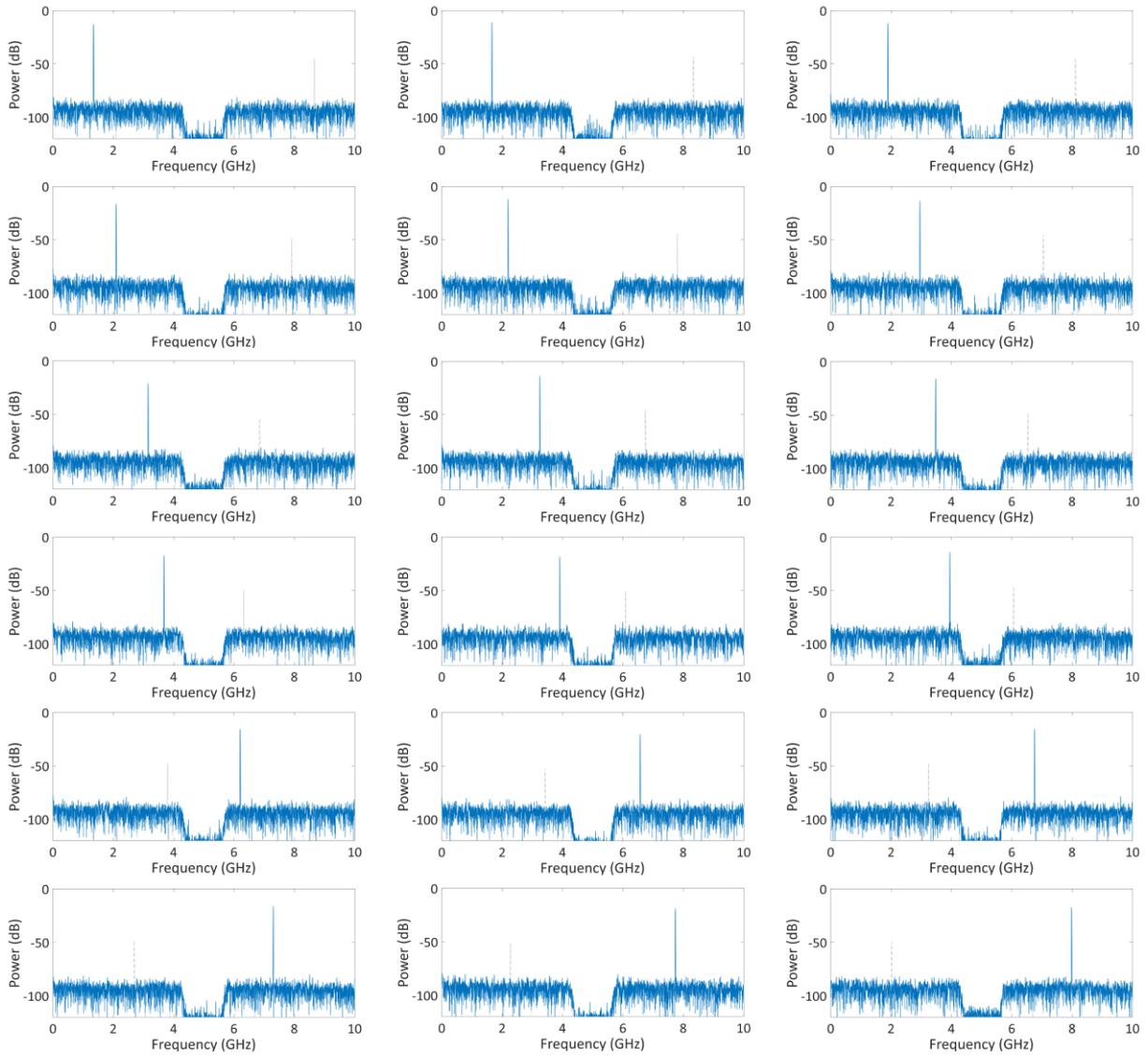

**Fig. S4. Sine signals results before and after the matching nets.** These data are randomly chosen in the validation set. Grey dashed curve is the mismatched interleaved data. Blue solid curve is the matching nets recovered data.

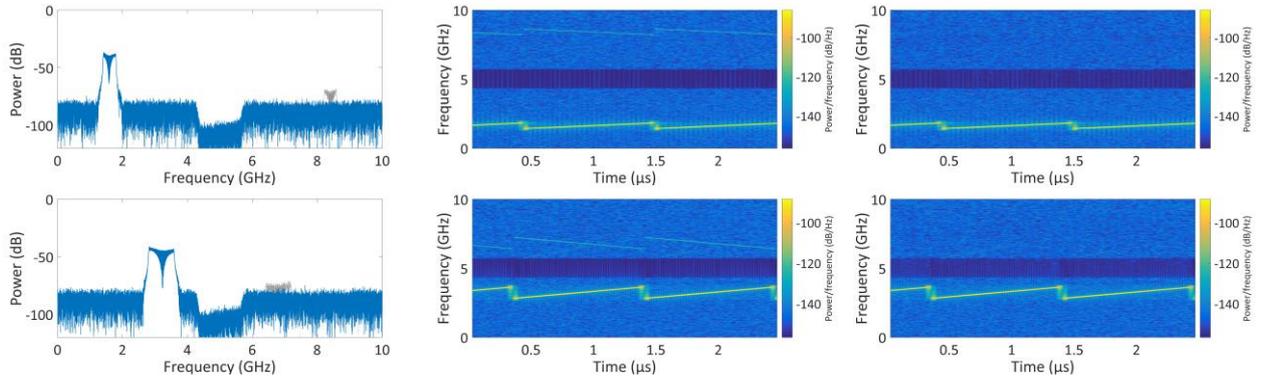

**Fig. S5. LFM signals results before and after the matching nets.** In all frequency spectra, grey dashed curves represent data before the matching nets and blue solid curves denote the recovered data after the matching nets. STFT plots are also given for LFM signals: the left and right columns are data before and after the matching nets, respectively.